\documentclass{JHEP3}
\usepackage{amsmath}
\usepackage{cite}
\usepackage{epsfig}
\usepackage{graphicx}

\def\be{\begin{equation}}
\def\ee{\end{equation}}
\def\baray{\begin{eqnarray}}
\def\earay{\end{eqnarray}}
\def\ba{\begin{eqnarray}}
\def\ea{\end{eqnarray}}

\title{Exact spectrum of scalar field perturbations in a radiation 
deformed closed de Sitter universe}

\author{Sash Sarangi $^1$, Koenraad Schalm $^2$, Gary Shiu $^3$, Jan Pieter
van der Schaar $^2$\\
$^1$ Institute of Strings, Cosmology and Astroparticle Physics,
Department of Physics, Columbia University, New York, NY 10027, USA\\
$^2$ Institute for Theoretical Physics, University of Amsterdam,\\
Amsterdam 1018 XE, The Netherlands \\
$^3$  Department of Physics, University of Wisconsin, Madison,
WI 53706, USA \\}
\date{\today}
\abstract{We observe that the equation of motion for a free scalar
field in a closed universe with radiation and a positive cosmological
constant is given by Lam\'e's equation. Computing the exact power
spectrum of scalar field perturbations, the presence of both curvature and 
radiation produces a red tilt weakly dependent on the amount of radiation.
}

                                         
\preprint{CU-TP-1164\\
MAD-TH-06-13\\
ITFA-06-45}

\begin{document}

\section{Introduction}
There is by now strong and growing evidence 
that the very early universe 
went through a phase of inflation \cite{Peiris:2003ff}. 
What led to this phase of 
accelerating expansion,
and what the underlying microscopic theory is, 
are questions that are still under intense investigation. 
Our current understanding from 
the nearly scale invariant spectrum of the cosmic microwave background (CMB)
anisotropies is that such an inflationary phase (at least the observable 
e-folds) can be well described by an approximate de Sitter universe. 

An interesting modification of a (quasi-) de Sitter spacetime that would
still be consistent with an inflationary history of the universe is
the introduction of some primordial radiation. At late times, this
radiation would 
get inflated away and the naive expectation is that this radiation will
therefore not leave any observable effects on the CMB temperature
anisotropy power spectrum. 
Various cosmological models have been proposed in which it is natural
to include 
some amount of radiation in the early (quasi-) de Sitter phase \cite{Berera:1995ie,McGreevy:2005ci,Danielsson:2005cc} and  
some general results on radiation-filled FRW universes can be found in  
\cite{Bouhmadi-Lopez:2002qz}.
Moreover, going beyond the mini-superspace approximation in selecting
the wave function of the universe \cite{Hartle:1983ai,Vilenkin:1982de,Linde:1984mx}
naturally leads to an effective description in terms of a radiation 
deformed de Sitter geometry, where the radiation density is related 
to the fundamental (string) length scale in the problem \cite{Sarangi:
2006eb,Sarangi:2005cs}. \footnote{A similar conclusion was reached in 
\cite{Brustein:2005yn} and \cite{Barvinsky:2006uh} based on string 
thermodynamic arguments and quantum gravitational loop effects, respectively.}

All these considerations warrant a more detailed study of a radiation deformed de Sitter geometry. 
In particular, it would be interesting to see whether 
primordial radiation could 
leave some observational signatures. 
Here we report that in
a de Sitter plus radiation universe, remarkably, 
the power spectrum of scalar field fluctuations is determined by a known
differential equation: Lam\'{e}'s equation. This allows for a precise analysis
of the spectrum and the modifications with respect to a pure de Sitter 
geometry.

Scalar fields in pure and deformed de Sitter spacetime have of course been
well studied. \cite{Mottola:1984ar,Allen:1985ux} studied the mode
functions of a scalar field on a pure de Sitter background.
The {\it open} radiation deformed de Sitter geometry 
is considered in \cite{Vilenkin:1982wt}. 
We study a closed radiation deformed de Sitter geometry. 
There is a crucial difference between the open and the closed geometries.
In an open radiation deformed geometry, 
the universe begins in a singularity, followed by a phase of radiation
domination, after which the universe becomes dominated by the
cosmological constant. In contrast, 
the radiation deformed {\it global} de Sitter spacetime interpolates between 
a pure closed de Sitter and an 
Einstein static universe as the total amount of radiation is varied from zero 
to the maximally allowed value, and there is no singularity in the geometry. 
The presence of any more radiation (than the maximum allowed amount)
leads to a spacelike singularity.
In Euclidean signature, the geometry
corresponds to a 
deformation of an $S^4$ to a ``barrel'' shaped geometry
\cite{Sarangi:2006eb,Sarangi:2005cs}. 
So below the maximally allowed value for the radiation
density the initial singularity is absent and the geometry can safely be analytically continued 
to a bouncing Lorentzian geometry. 

Our study is motivated in part because the Euclidean geometry allows
an interpretation as 
the instanton process involved in selecting the
wave function of the universe
\cite{Hartle:1983ai,Vilenkin:1982de,Linde:1984mx}. 
What might be a natural choice for the wave function of the universe is
a subject of long debate \cite{Vilenkin:1998rp}, 
which we will not address in this paper (see \cite{Sarangi:2006eb,
Sarangi:2005cs} for discussion). 
In the original Hartle-Hawking context, it allows for an
unambiguous determination of the vacuum state of fluctuations
\cite{Halliwell:1984eu,Vachaspati:1989wf}. The cosmological vacuum ambiguity has recently received renewed interest
as future CMB experiments are sensitive enough to probe it 
\cite{Greene:2005aj}. We will
point out that in a Hartle-Hawking cosmological scenario {\em with}
radiation the Bunch-Davies initial state is no longer unambiguously selected by analytic continuation to the Euclidean geometry.

To briefly summarize our main result: the effect of the radiation is
to lead to a moderation of the closed dS red tilt of the scalar field 
power spectrum. 
The obvious explanation for this is that the low 
$k$-modes are 
sensitive to the spatial curvature, i.e. the compactness, of the closed 
de Sitter geometry. 
The 
radiation deformation 
effectively
shrinks the volume of the spatial $S^3$'s, 
which changes
the red tilt in the fluctuations. 
Consistency with the observed flatness of our universe, however, make
the red-tilt due to curvature 
barely observable.
Similarly, 
the additional red-tilt due a radiation deformation is unlikely to
show up in future experiments. 

\section{A closed universe with a positive cosmological constant and radiation}
The scale factor for
a closed universe with a positive cosmological constant
$\Lambda$ plus a radiation term with energy density $ \rho_*/a^4$
obeys the equation
\baray 
\label{eq:1bc}
2\frac{1}{a(t)} \frac{d^2a(t)}{dt^2}  +
\frac{1}{a(t)^2} \left( \frac{da(t)}{dt} \right)^2 +
\frac{1}{a(t)^2} = \Lambda - \frac{\rho_*}{a(t)^4}~, 
\earay 
We use $t$ for time in Lorentzian signature metric, $\tau$ for time
in Euclidean signature metric, natural units in which $8\pi
G=c=\hbar=1$. 
The solution to Eqn. (\ref{eq:1bc}) is given by 
\baray 
\label{barrel1} 
a(t) = \frac{1}{\sqrt{2}L} \sqrt{\left( 1 + \sqrt{1 - \beta} 
\cosh (2Lt) \right)}~. 
\earay 
Here $\beta = 4L^2 \rho_*$ and $L=\sqrt{\frac{1}{3} \Lambda}$ is the
late-time horizon (see appendix \ref{sec:appendix-b-}). The
Euclidean version is a deformation of $S^4$. We will refer to it as the
``barrel'' geometry \cite{Sarangi:2005cs} 
\baray 
\label{barrel2}
a(\tau) = \frac{1}{\sqrt{2}L} \sqrt{\left( 1 + \sqrt{1 - \beta} \cos
(2L\tau) \right)}~. 
\earay 
As $\beta \to 0$ (no radiation) one
recovers the pure de Sitter geometry. In the opposite limit $\beta \to
1$ one obtains the Einstein Static universe. The Euclidean geometry
corresponds to an $S^4$ for $\beta=0$, a cylinder ($R^1 \times S^3$)
for $\beta=1$, and for general $\beta$ interpolates between
these two, resembling a barrel (Fig.($1$)).\\

Perhaps a geometrically more intuitive way to write the scale factor
is 
\baray 
\label{rad} a(\tau) =
\frac{1}{L\sqrt{1+2\Delta^2}}\sqrt{\Delta^2  + \cos^2 (L \tau)}~.
\earay 
where $\Delta^2 = \frac{1-\sqrt{1-\beta}}{2\sqrt{1-\beta}}$.
In the limit of vanishing radiation, $\Delta \to 0$, one manifestly
recovers pure de Sitter.

Compared to Lorentzian de Sitter (Eq.(\ref{barrel1})), the size of the spatial $S^3$'s shrink 
as a consequence of the radiation, which is most pronounced around the 
waist at $t=0$. At this bounce, the waist 
is reduced by a factor $\sqrt{\frac{1+\Delta^2}{1+2\Delta^2}}$. 
This effect reaches a minimum in the (critical) limit 
$\beta \rightarrow 1$, or equivalently $\Delta^2 \rightarrow \infty$. 
The geometry reduces to a pure de Sitter spacetime as $t \rightarrow
\pm \infty$ due to the redshifting of the radiation.
The Penrose diagram of the radiation deformed solution 
corresponds to a slightly elongated version of the perfect square of
pure de Sitter \cite{Leblond}. This shows that the compact nature of the 
spatial slices can in principle be ascertained in a finite amount of proper 
time by fiducial observers, which is impossible in pure de Sitter.

\begin{figure}
\begin{center}
\includegraphics[width=9cm]{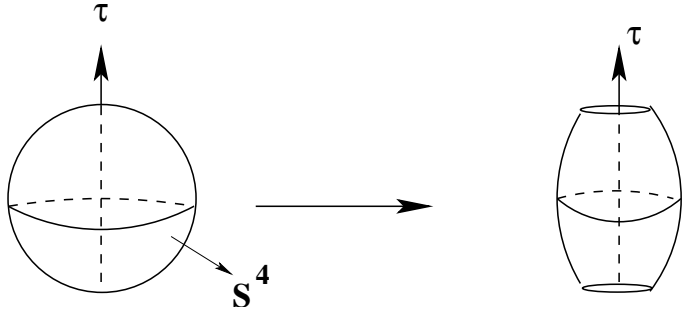}
\vspace{0.1in}
\caption{Radiation changes the Euclidean $S^4$ to a Euclidean Barrel.}
\label{fig1}
\end{center}
\end{figure}

\begin{figure}
\begin{center}
\includegraphics[width=9cm]{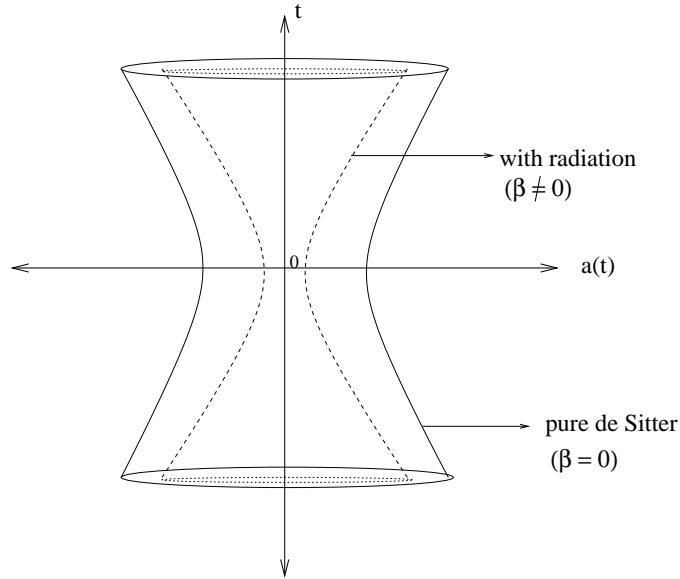}
\vspace{0.1in}
\caption{The solid figure corresponds to the pure de Sitter universe
(Euclidean $S^4$). The dotted figure corresponds to a radiation deformed de Sitter universe
(Euclidean Barrel). More radiation leads to more shrinking at the waist.}
\label{fig2}
\end{center}
\end{figure}

\section{Scalar field in a closed de Sitter universe with radiation}

To show that the fluctuation spectrum is determined by the Lam\'{e}
equation, 
consider a minimally coupled scalar field 
\baray
{\cal L}_{\Phi} = -\frac{1}{2}g^{ab}(\partial_a \Phi)(\partial_b \Phi)
- \frac{M^2}{2}\Phi^2 ~.
\earay
The wave equation (in Euclidean signature) is 
given by
\baray
\left(-\frac{\partial^2}{\partial \tau^2} - 3\frac{\dot{a}(\tau)}{a(\tau)}
 \frac{\partial}{\partial \tau} - \frac{\Delta_3}{a(\tau)^2} + M^2\right)
\Phi = 0~,
\earay
where $\dot{a} \equiv d a(\tau)/d\tau$, 
and
$\Delta_3$ is the Laplacian on the unit three-sphere.
As is standard, 
we decompose the field into $S^3$ spherical harmonics 
\baray
\Phi = f_k(\tau) Y_{klm}(\chi, \theta, \phi)
\earay
with eigenvalues
\baray
\label{eigenvalues}
\Delta_3 Y_{klm} = - (k-1)(k+1) Y_{klm} \, ~~~~ (k-1) \geq l \geq |m|
\geq 0. 
\earay
The mode function $f_k(\tau)$ must then obey the following equation
\baray 
\label{f_eq} 
- \ddot{f_k}- 3 \frac{\dot{a}}{a}\dot{f_k}+
\left[ \frac{(k^2-1)}{a^2} + M^2 \right] f_k = 0 \, . 
\earay
We are following the common convention in a closed cosmology to let the 
minimum (co-moving) momentum variable $k$ equal $1$, corresponding to the
homogeneous mode of Eq.(\ref{eigenvalues}). 
Note that
in our notation 
the scale factor $a(\tau)$ has dimensions of length, implying that
the co-moving momentum $k \equiv a(\tau) \, p$ is a dimensionless
number. An important thing to realize is that in these conventions the same
co-moving scale in initially different closed universes (for example one with or 
without primordial radiation), will not correspond to the same physical scale at 
late 
cosmological time. 

In what follows, we shall simplify the above equation and show that
the mode functions $f_k(\tau)$ are given in terms of solutions to
Lam\'{e}'s equation. First make the
substitution 
\baray 
F_k(\tau) = a(\tau) f_k(\tau)~. 
\earay 
then $F_k$ must satisfy 
\baray 
\ddot{F_k} + \frac{\dot{a}}{a}\dot{F_k} -
\left( \frac{\ddot{a}}{a}+\frac{\dot {a}^2}{a^2} \right)F_k -
\left(\frac{(k^2-1)}{a^2} + M^2 \right)F_k = 0~. 
\earay
  From the trace of the Einstein equations it is easy to show that for a 
closed FRW universe (in Euclidean signature)
\baray 
\label{eq:2} 
\left( \frac{\ddot{a}}{a}+\frac{\dot {a}^2}{a^2}
\right) = -2 L^2 + \frac{1}{a^2}~. 
\earay 
This leads to the equation
\baray 
\label{eqn} 
\ddot{F_k} + \frac{\dot{a}}{a}\dot{F_k}+ L^2
\left( \mu(\mu + 1) - \frac{k^2}{L^2 a^2}
 \right) F_k = 0~,
\earay 
where $\mu(\mu +1) = 2 - M^2/L^2$. Substituting the explicit
form of the scale factor,
we obtain the following equation
\baray
\left(1+ \frac{\Delta^2}{\sin^2(L \tilde{\tau})} \right) \ddot{F_k} + L \cot
(L\tilde{\tau}) \dot{F_k} + L^2 \left( \mu(\mu + 1)+ \frac{\Delta^2(\mu
(\mu + 1)) - k^2(1+2\Delta^2)}{\sin^2(L\tilde{\tau})} \right)F_k =0 \nonumber \\
\earay in terms of $\tilde{\tau} = \tau + \pi/2L$.  In the limit
$\Delta \to 0$, one recovers the associated Legendre equation for
pure de Sitter space using global coordinates (closed $S^3$ slicing)
\cite{Starobinsky:1996ek, Chernikov:1968zm}.\\

Let us now express everything in terms of a variable $y = \cos^2(L\tilde
{\tau})$. This yields
\baray
\label{heun}
F''_k(y) + \frac{1}{2}\left(\frac{1}{y} + \frac{1}{y-1}+\frac{1}{y-z}
\right)F'_k(y) & \nonumber \\
+ \frac{ (\mu(\mu +1) - k^2 ) + \Delta^2(\mu(\mu +1) - 2 k^2)-
\mu(\mu +1)y }{4y(y-1)(y-z)} F_k(y) = 0~, \earay where $ z = 1 +
\Delta^2$. Eq.(\ref{heun}) is an ordinary differential equation with
four singular points (at $0, 1, z, \infty$). This equation is well
studied and is known as Heun's equation \cite{Heun:1995hn}. In fact,
Eq.(\ref{heun}) is a special type of Heun's equation called Lam\'{e}´s
equation. That scalar field fluctuations in closed de Sitter
universes plus radiation reduce to Lam\'e's equation was also noted
in \cite{Bouhmadi-Lopez:2002qz}.  Here we use this to compute an
exact power-spectrum of scalar field fluctuations. In this context
it is interesting to note that according to \cite{Heun:1995hn},
Lam\'e's equation is the most general equation of Heun type which
can arise from a Laplacian after separation of variables. For the
sake of completeness the general features of Heun's and Lam\'e's
equation are discussed in appendix \ref{sec:appendix-a:-heuns}.
Solutions to Lam\'e's equation exist both as a power series and as a 
series in hypergeometric functions.

\subsection{Initial conditions and vacuum states}

To determine the complete solution to the mode equation (\ref{f_eq})
we must supply initial conditions or equivalently define the vacuum
state. 
This is where physics enters.
To construct the vacuum states we follow
\cite{Allen:1985ux,Allen:1987tz}. In the previous sections we wrote 
the scalar field equation of motion in a metric using
Euclidean time. In this section we will work in Lorentzian
signature, 
$a(t) = \frac{1}{L \,\sqrt{1+2\Delta^2}} \sqrt{\Delta^2 + \cosh^2(L t) } $.

In order to follow the notation of \cite{Allen:1987tz}, we start from the Lorentzian
counterpart of Eq.(\ref{eqn}), and define a new time variable $\eta$ given by
\baray 
\label{eq:33b}
\eta = 2\arctan\left( e^{L t} \right)~. 
\earay 
As $-\infty < t < \infty$, $ 0 < \eta < \pi$. For pure de Sitter this variable 
simply corresponds to conformal time, but this is not true in general. 
In terms of this variable the equation of motion becomes
\baray 
\label{eq:33c}
\frac{d^2F_k}{d\eta^2} + \frac{\Delta^2 \sin\eta \cos\eta}{1
+ \Delta^2 \sin^2\eta} \frac{dF_k}{d\eta} + \left(
\frac{(1+2\Delta^2) k^2}{1+\Delta^2 \sin^2 \eta} - \frac{(2
- M^2/L^2)}{\sin^2\eta} \right)F_k = 0~. 
\earay
It will be useful to recast this equation one more time into a
deformation of the defining associated Legendre equation by
introducing a new function $G_k$ related to $F_k$ as $F_k = \left(
\sqrt {\sin\eta}/ (1+\Delta^2 \sin^2\eta)^{1/4} \right) G_k$. Then
in terms of the variable $x = -\cos(\eta)$ one obtains 
\baray 
\label{mod} 
(1 - x^2)\frac{d^2G_k(x)}{dx^2} - 2x \frac{dG_k(x)}{dx}
+ \frac{1}{1+ \Delta^2(1-x^2)} \left( (1+2\Delta^2)(k^2 - 1/4)-
\right. \nonumber \\ \left. \frac{v^2(1+\Delta^2(1-x^2))}{1-x^2} +
\frac{3\Delta^2(1+\Delta^2)}{4}\frac{(1-x^2)} {(1 +
\Delta^2(1-x^2))} \right)
G_k(x) = 0 ~,\nonumber \\
\earay
where $v^2 = 9/4 - M^2/L^2$. \\

It is worth briefly recapitulating the construction of vacua for de
Sitter spacetime before we complicate matters by taking into account
the radiation term. When $\Delta^2 = 0$, the above equation is by
construction an associated Legendre equation, and the time variable 
$\eta$ corresponds to conformal time.
The most general solution is a linear combination of the associated Legendre polynomials
\baray
\label{DS}
F_k(\eta) = \sqrt{\sin\eta}[A_k P^v_{k-1/2}(-\cos\eta) + B_k Q^v_{k-1/2}
(-\cos\eta)]~.
\earay
The coefficients $A_k$ and $B_k$ are not independent. The orthonormality of 
the physical mode functions requires that
\baray
\label{constrain}
F_k\frac{dF^*_k}{d\eta} - F^*_k \frac{dF_k}{d\eta} = i~,
\earay
and this leads to the constraint
\baray
\label{alpha}
A_kB_k^* - B_k A^*_k = i \frac{\Gamma(k+1/2-v)}{\Gamma(k+1/2+v)}.
\earay

A priori there is no preferred vacuum state and all solutions to
equation (\ref{alpha}) are allowed. On physical grounds however,
one state, the Bunch-Davies or Euclidean vacuum is usually preferred. 
The Euclidean vacuum is the unique set of initial conditions for which
the analytically continued 
Euclidean Green's function has no singularities at the antipodal
point \cite{Strominger}.  For our purposes this can simply be interpreted to mean 
that the state should reduce to the usual flat space vacuum at 
physical momentum scales much larger than the curvature scale, i.e. $k/a \gg H$. This requirement singles out the following values for the coefficients
\cite{Allen:1985ux, Allen:1987tz}
\baray
A_k &=& \left( \frac{\pi}{4} \frac{\Gamma(k+1/2-v)}{\Gamma(k+1/2+v)}
\right)^{1/2} e^{i\pi v/2}, \nonumber \\
B_k &=& \frac{-2i}{\pi}A_k.
\earay
In the 
Hartle-Hawking ``tunneling from nothing'' scenario for the wavefunction
of the universe, analyticity has been argued to uniquely resolve the 
cosmological vacuum ambiguity 
in favor of this Euclidean Bunch-Davies vacuum
\cite{Halliwell:1984eu,Vachaspati:1989wf}. 
Clearly this analyticity argument no longer holds  with even a small 
amount of radiation present; there is no regularity condition at the 
south pole of the ``barrel''.

\subsection{Vacua for the radiation deformed geometry}
We would like to see how this de Sitter result - the one complex
parameter family of vacua given by Eq.(\ref{alpha}) - gets modified by
the presence of radiation. We do so by a power series analysis of
solutions to Eq.(\ref{mod}). Let us denote the two independent solutions by
$G^v_{k-1/2}$ and $H^v_{k-1/2}$. We will require that in the limit of vanishing radiation
$G^v_{k-1/2}$ reduces to $P^v_{k-1/2}(x)$ and $H^v_{k-1/2}$ reduces to $Q^v_{k-1/2}$.
Using the power series ansatz $G(x) = \sum_{r=0}^{\infty} c_r (1-x)^{r + \alpha}$
it is straightforward to verify that the two independent solutions to Eq.(\ref{mod}) 
are given by
\baray
\label{result}
\alpha &=&  v/2,~~~~~ \mbox{or} ~~\alpha = -v/2  ~,\nonumber \\
 \frac{c_1}{c_0} &=&  \frac{2\alpha(2\alpha +1) - 2(k-1/2)(k+1/2)}
{(2\alpha +2)^2 -v^2} + 4\Delta^2 \frac{v^2 - 4\alpha^2
-(k-1/2)(k+1/2)}{(2\alpha + 2)^2 -v^2}~. 
\earay
Here we have listed only the ratio of the first two coefficients. This suffices
to study the late time behavior ($x \to 1$ limit). The coefficient
$c_0$ is fixed by requiring the solutions to reduce to the associated Legendre
functions for zero radiation. This gives $c_0 = \left( (-1)^v \Gamma(k+v+1/2) \right) /
\left( 2^{v/2} v! \Gamma(k-v+1/2) \right)$.  The general $n$-th term
recursion relation is easy to find.

The above two values of $\alpha$ (viz. $\alpha = +v/2, -v/2$) give
two independent solutions $G^v_{k-1/2}$ and $G^{-v}_{k-1/2}$. These
are the analogs of $P^v_{k-1/2}$ and $P^{-v}_{k-1/2}$, respectively.
To find $H^v_{k-1/2}$ - the analog of $Q^v_{k-1/2}$ - we remind the
reader of the relationship for associated Legendre functions
\cite{GR2000} 
\baray 
\label{PQ} 
Q^v_{k-1/2}(x) =
\frac{\pi}{2\sin(v\pi)} \left(P^v_{k-1/2}(x) \cos(v\pi)
-\frac{\Gamma(k+v+1/2)}{\Gamma(k-v+1/2)} P^{-v}_{k-1/2}(x)\right)~.
\earay 
We can find the corresponding relation for $H^v_{k-1/2}$ as
follows. In Eq.(\ref{PQ}), the coefficients of the associated
Legendre functions are independent of the variable $x$, representing
time. We expect the same to hold true for the modified relationship.
If we can fix these coefficients at one instant of time, they will
have those values at all times. Now note that at late times (as $x
\to 1$), Eq(\ref{mod}) takes the form 
\baray 
\label{mod'} (1 -
x^2)\frac{d^2G_k(x)}{dx^2} - 2x \frac{dG_k(x)}{dx} +
\left((1+2\Delta^2)(k^2 - 1/4) - \frac{v^2}{1-x^2} \right)
G_k(x) = 0~.  \nonumber \\
\earay 
This is almost the equation for de Sitter spacetime, except
for the factor of $(1+2\Delta^2)$. This might seem puzzling. At late
times one expects the radiation to have no effect. However, recall
that even at (fixed) late times, the overall size of a compact spatial 
slice is scaled down due to the presence of the radiation. 
Now we can construct a relationship analogous to Eq.(\ref{PQ}). At
late times, $G^v_{k-1/2}$ approaches $P^v_{\chi-1/2}$, with
$(\chi^2-1/4)= (1+2\Delta^2)(k^2-1/4)$. This suggests to simply
replace $k$ by $\chi$, giving 
\baray 
\label{GH} H^v_{k-1/2}(x) =
\frac{\pi}{2\sin(v\pi)} \left(G^v_{k-1/2}(x) \cos(v\pi)
-\frac{\Gamma(\chi+v+1/2)}{\Gamma(\chi-v+1/2)}
G^{-v}_{k-1/2}(x)\right)~. 
\earay

In terms of these two independent solutions $G^v_{k-1/2}$ and $H^v_{k-1/2}$, the general
solution to the mode functions reads
\baray
\label{eq:DSpR}
F_k &=& \frac{\sqrt(\sin \eta)}{(1+\Delta^2 \sin^2\eta)^{1/4}} G_k \\ \nonumber
&=& \frac{\sqrt{\sin \eta}}{(1+\Delta^2 \sin^2\eta)^{1/4}}
[\tilde{A_k} G^v_{k-1/2}(-\cos \eta) +
\tilde{B_k} H^v_{k-1/2}(-\cos \eta)  ]~.
\earay
A careful analysis of the orthonormality constraints shows that 
the two complex parameters $\tilde{A_k}$ and $\tilde{B_k}$ must be related as
\baray 
\label{alpha'} 
\tilde{A}_k \tilde{B}_k^* - \tilde{B}_k
\tilde{A}^*_k = i \, \sqrt{1+2\Delta^2} \,
\frac{\Gamma(\chi+1/2-v)}{\Gamma(\chi+1/2+v)}.
\earay
The appearance of another $\sqrt{1+2\Delta^2}$ factor in this condition is important. 
It appears because 
in the radiation deformed geometry the $\eta$ variable 
does not correspond to conformal time. From 
Eq.(\ref{rad}) 
one 
sees that in the late time limit, the scale factor reduces to 
$a(t) \approx \frac{\cosh{Lt}}{L\sqrt{1+2\Delta^2}}$. As compared to 
pure de Sitter space, this simply 
amounts
to a rescaling of the conformal time variable 
with $\sqrt{1+2\Delta^2}$. This explains the factor 
in Eq.(\ref{alpha'}). 
In addition we simply replaced $k$ by $\chi$ on the right-hand-side of
this equation, following our earlier argument that at late times the Eq.(\ref{mod}) 
reduces to the associated Legendre equation and that the normalization
condition is independent of time.

To find the analog of the Bunch-Davies 
vacuum in the (late time) 
radiation deformed geometry, we proceed in exactly the same way as before. 
We 
demand that our solution reduces to the usual flat space vacuum 
at high momentum $p=k/a$. In addition it should reduce to the
usual de Sitter Bunch-Davies vacuum in the limit $\Delta^2 \rightarrow 0$.
Keeping in mind that $(\chi^2-1/4)=(1+2\Delta^2)(k^2-1/4)$, this
singles out the following solution
\baray
\label{BD'}
\tilde{A_k}^{BD} &=& \left( 1+2\Delta^2 \right)^{1/4} \, \left( \frac{\pi}{4} \frac{\Gamma(\chi+1/2-v)}{\Gamma(\chi+
1/2+v)}\right)^{1/2} e^{i\pi v/2}, \nonumber \\
\tilde{B_k}^{BD} &=& \frac{-2i}{\pi}\tilde{A_k}^{BD}~.
\earay
So we see that the only difference as compared to the pure de Sitter
case is the replacement of $k$ with $\chi$ and the additional factor
of $\left( 1+2\Delta^2 \right)^{1/4}$ due to the change in the normalization 
condition Eq.(\ref{alpha'}).

Fundamentally there is deeper difference between the Bunch-Davies in
the radiation deformed case and that of the pure de Sitter. Were one
to insist on analytic behavior of the wave function in the Euclidean
past, as e.g. in the Hartle-Hawking scenario, then one must choose the
Euclidean or Bunch Davies vacuum in pure de Sitter
\cite{Halliwell:1984eu,Vachaspati:1989wf} to avoid a singularity at
the south pole of the sphere. Because even a tiny amount of radiation
opens up the South pole to a barrel, there appears to be no reason to
insist on well-defined behavior at the edge. The Bunch-Davies state is
now only preferred for the usual reasons. 

Finally note that one should no longer consider the limit $\Delta^2
\rightarrow \infty$ in the late-time de Sitter solution; 
the late-time de Sitter limit is implicitly defined as  $t \geq t_f$, 
with  $\cosh^2{(L t_f)} \gg \Delta^2$.
Instead we know that the homogeneous solution reduces to the Einstein 
Static Universe in the 
limit $\Delta^2 \rightarrow \infty$.
It is
not difficult to 
see that in that case the mode-functions will reduce to 
those of ordinary flat space, i.e. plane waves. As we will be interested
in a (late-time) inflationary stage, the late-time de Sitter solution
is the appropriate one to study the leading effects due to the 
radiation deformation.

\section{The power spectrum of scalar field perturbations}
The large scale density perturbations present in the CMBR temperature
anisotropy we see today 
crossed the horizon 
about $55$ 
e-folds before the end of the inflation.\footnote{The precise e-folds at which the observable modes crossed the horizon depend on the reheating temperature. In some extreme cases, the horizon crossing can be as late as $25$ e-folds before the end of inflation.} Subsequently they remained
frozen till horizon re-entry. Assuming we can construct a suitable
closed inflationary model based on the radiation deformed de Sitter 
geometry, 
the question whether 
there will be any
imprint 
on these perturbations
probably depends on whether the radiation was a sizeable
component at horizon exit. So we expect that the large
wavelength/small $k$ 
perturbations 
are more likely to be
affected by the radiation (as they cross the horizon and get frozen
in early on during inflation).

Let us remind the reader that we will calculate the power spectrum of massless
scalar field perturbations, which
is 
not the same as the primordial power spectrum of scalar density perturbations.
Nevertheless, as is usual, we will implicitly assume that parts of our results 
continue to hold for the primordial spectrum of scalar density 
perturbations in a slow-roll inflationary generalization of the background 
geometry, which is the 
relevant quantity responsible for the observable CMBR anisotropy spectrum.

The advantage of having an exact (late-time) solution to the equations of 
motion is that 
the power spectrum of fluctuations is precisely given by
\begin{equation}
P_{\phi} = \lim_{t\rightarrow \infty}^{\rm leading~term} \frac{k^3}{2\pi^2}|f_k(t)|^2 \, ,
\end{equation}
where the correctly normalized mode function for a massless field
($v=3/2$) equals
 \baray
f_k = \frac{L\sqrt{1+2\Delta^2}(\sin \eta)^{3/2}}{(1 + \Delta^2 \sin^2 \eta)^{3/4}}
[\tilde{A}_k G^{3/2}_{k-1/2} + \tilde{B}_k H^{3/2}_{k-1/2}] .
\earay
In the late time limit, just like in the pure de Sitter case, the
second term will have vanishing contributions, $H^{3/2}_{k-1/2} \to 0$.
Using 
that in the late time limit $G^{3/2}_{k-1/2}$ approaches the
associated Legendre polynomial $P^{3/2}_{\chi-1/2}(x)$, with asymptotic behavior
\begin{equation}
\lim_{x\rightarrow 1} P^v_m (x) = (1-x)^{-v/2} \frac{2^{v/2}}{\Gamma(1-v)} \, ,
\end{equation}
the late time mode function is
\baray |f_k|^2
=\left( 1+2\Delta^2 \right)^{3/2} \frac{L^2}{2 (\chi^2-1)\chi} + \ldots~. 
\earay 
To evaluate $\chi$, recall that $(\chi^2 - 1/4) = (1+2\Delta^2)(k^2-1/4)$.
Solving for $\chi$ we get 
\baray 
\chi = k \sqrt{1 + 2\Delta^2
\frac{(k^2-1/4)}{k^2}}~. 
\earay 
Thus the {\em exact} scalar field power spectrum is
given by
\baray 
\label{power1} 
P_\Phi =
\frac{k^3}{2\pi^2} |f_k|^2
&=& \left( 1+2\Delta^2 \right)^{3/2} \, \frac{L^2}{4\pi^2} \, 
\frac{k^3}{(\chi^2-1)\chi} ~.
\earay
This is our main result. For large wavenumber $k$ it reduces to
\baray
\label{power} 
P_{\Phi} &=& \frac{L^2}{4\pi^2}\left(
 1 + \frac{1}{k^2} \left[ \frac{1+ \frac{3}{4} \Delta^2}{1+2\Delta^2}
 \right] + \ldots \right)~. 
\earay
The overall $\frac{L^2}{4\pi^2}$ term 
corresponds to the well-known de Sitter 
result. 
Notice 
the cancelation of the radiation dependent 
factors between the norm of the mode-function and the explicit 
$\chi$ dependence. 
The term in between brackets corresponds to a small red-tilt 
(more power at small $k$) of the spectrum of scalar field perturbations. At low values 
of $k$, the departure from scale invariance is pronounced and 
the radiation component is of the same order (in $1/k$) 
as the effect due to curvature alone \cite{Starobinsky:1996ek}. 
The red tilt is of course 
easy to understand. Low $k$-values correspond to those modes which exited the 
horizon early on and were, therefore, more sensitive to the 
radiation present 
and the compact nature of the spatial $S^3$ slices. At large values of $k$, 
the radiation as well as the spatial curvature has inflated away and the 
corresponding modes would fail to notice any sizeable effect due to radiation 
or curvature. 

\def\Ome{\Omega}
With the exact power spectrum in hand, the next question is naturally
whether the deviations from pure de Sitter are observable. 
Obviously in short inflation scenarios there is a better chance of 
detecting such an effect. 
Recall that in closed universes there is a minimal wavenumber
corresponding to the size of the spatial $S^3$. The longer inflation
lasts, the higher the wavenumbers are that are responsible for the
observed CMB, and thus the smaller the effect. Inflation must have lasted long enough to solve the
horizon and flatness problems, however, and observability thus hinges
on the minimal length allowed by these constraints. Flatness in
particular constrains the present scale factor $a_p$ in terms of the
observed relative density
\begin{eqnarray}
  \label{eq:18}
  \Ome_{tot}(t) = 1+ \frac{1}{a^2H^2} ~~~~~& \rightarrow &~~~~~ a_p^2H_p^2\, \geq\, \frac{1}{\Ome_{obs}-1}~.
\end{eqnarray}
The largest scale visible in the CMB,
$k=a_{p}H_{p}$, 
must therefore be equal to or larger than
\begin{eqnarray}
  \label{eq:17}
  k^2 & \stackrel{>}{=} &a^2_{p}H^2_{p} =
  \frac{1}{(\Ome_{obs}-1)}
\end{eqnarray}
Using the observed value $\Ome_{obs} \simeq 1.02$, the maximal
correction to the power spectrum is therefore of order $10^{-2}$. A
red tilted correction of this size is in
practical terms barely distinguishable due to cosmic variance.

\section{Discussion and Conclusion}
To summarize, a radiation deformed inflationary phase is an interesting
extension to the current minimal paradigm.
A small amount of radiation is in fact quite generic.
Surprisingly, the radiation deformed 
geometry allows for an exact analysis of probe scalar field perturbations. 
Calculating the power spectrum of scalar field fluctuations we found that 
the relative power at low multipoles is enhanced. Both the curvature
and the radiation component contribute at the same order and together
give rise to a red 
power spectrum. 
The observed near critical density of the universe,
however, implies 
that the magnitude of this effect 
is barely detectable. 
Finally, in the context of a Hartle-Hawking ``tunneling-from-nothing''
scenario, the small amount of radiation expected to be present
\cite{Sarangi:2006eb,Sarangi:2005cs,Firouzjahi:2004mx} 
implies that the Euclideanized geometry - a barrel rather than a sphere -
does no longer single out a
unique vacuum state.
For the barrel there is no analyticity constraint that selects the
the Bunch-Davies initial state.

We conclude with a (partial) list containing some suggestions on how our
results might connect to other work.

\begin{itemize}
\item As pointed out, in (short) closed inflation scenarios, the radiation 
effects 
are most prevalent.
In the context of the string theory landscape of de Sitter vacua, 
a Hartle-Hawking ``tunneling from nothing'' origin of the universe might favor short inflation scenarios over 
inflationary phases with a large number of e-folds  because
it is far easier to arrange a few e-folds in the landscape.
Assuming a random form of the de Sitter potentials in the landscape,
a statistical analysis of the possibility of inflation
can be done as in \cite{Aazami:2005jf}. 
In fact,
the proposal in \cite{Arkani-Hamed:2005yv} ("living dangerously") 
suggests that the physical parameters of our
universe should be on the verge of violating important experimental bound
(favoring, for example, short inflation).

\item In \cite{McGreevy:2005ci} an initial state of the
universe is constructed from closed string tachyons. This
initial state corresponds to a thermal excitation above the
Bunch-Davies initial state. Clearly, due to backreaction of the
thermally excited state (i.e. radiation), the geometry will be
modified and one might expect the power spectrum to be affected 
in a similar way as discussed here.

\item In brane inflation scenarios, one expects a period
of collision and annihilation of branes prior to the last
phase of brane inflation directly preceding the reheating of our
universe. In such scenarios, again assuming that short inflation
is favored over long periods of inflation, we expect that
there will generically be a radiation component (due to brane
annihilation) during inflation.

\item It is interesting to compare our result with the warm inflation
scenario \cite{Berera:1995ie}. Warm inflation consists of a self
interacting inflaton field with a dissipative mechanism that
generates a sizeable radiation density during inflation. So
instead of having a reheating stage following a supercooled
inflationary phase, radiation is constantly generated during
inflation. As discussed in \cite{Berera:1995ie}, for certain
models of the inflaton potential and dissipation mechanisms
there can be a blue, instead of a red, tilt in the density 
perturbation for small wavenumbers due to thermal effects.  

\item The authors of \cite{Watson:2006px} consider a scenario where
the initial state of the universe is described by a positive cosmological
constant and radiation (motivated by the modified wave function discussed in
\cite{Sarangi:2006eb,Sarangi:2005cs,Firouzjahi:2004mx}). The subsequent 
evolution of the universe is quite interesting. The presence of a 
periodic potential (motivated by the string landscape) leads to 
a Bloch band for the wave function, and the universe cascades down this
band structure (having originated somewhere near the top of the band
as implied by the modified wave function)
as it evolves. In the process, the energy in the positive cosmological
constant gets converted into radiation until there is too little
energy in the cosmological constant to support an inflationary phase.
The initial state of a scalar field in this scenario will again be
described by the mode functions that we have found in this paper.
The subsequent evolution of the mode functions through the cascades
warrants a more detailed study.

\end{itemize}

\acknowledgments

We thank Xingang Chen, Pier-Stefano Corasaniti, Joel Erickson, Brian Greene,
Justin Khoury, Maulik Parikh, Henry Tye for useful discussions. We also thank
the Institute for Theoretical Physics Amsterdam and the
University of Wisconsin, Madison for hospitality. This work is supported in part by
the DOE under Grants No. DE-FG-02-92ER40699 (SS) and DE-FG-02-95ER40896(GS), 
NSF CAREER Award No. PHY-0348093 (GS), a Research Innovation Award (GS) and 
Cottrell Scholar Award (GS) from Research Corporation,  by Stichting FOM (JPvdS), 
and by a VIDI Award from the Netherlands Organization for Scientific Research (NWO)(KS).

\vspace{0.5cm}

\appendix

\section{Appendix: Heun's and Lam\'e's Differential Equation}
\label{sec:appendix-a:-heuns}

Heun's differential equation is of Fuchsian type with regular singularities
at $z = 0,1,a,\infty$.
The canonical form of Heun's differential equation is
\baray
\label{heuneqn}
\frac{d^2y}{dz^2} + \left( \frac{\gamma}{z} + \frac{\delta}{z-1} +
\frac{\epsilon}{z-a} \right) \frac{dy}{dz} + \frac{\alpha \beta  z -q}
{z(z-1)(z-a)} y =0
\earay
where $y$  and $z$ are regarded as complex variables and $\alpha$, $\beta$,
$\gamma$, $\delta$, $\epsilon$, $q$, $a$ are complex and arbitrary parameters
, and $a \neq 0, 1, \infty$. The first five parameters are linked by the equation
\baray
\label{reln}
\gamma + \delta + \epsilon = \alpha + \beta + 1
\earay
The exponents at the singular points $z = 0,1,a,\infty$ are
respectively $ (0, 1-\gamma); (0, 1-\delta); (0, 1-\epsilon);
(\alpha, \beta)$. According to the general theory of Fuchsian
equations the sum of these exponents must be equal to $2$. This
requirement leads to Eq.(\ref{reln}). It can be shown that any
Fuchsian second-order differential equation with four singularities
can be reduced to the form given by Eq.(\ref{heuneqn}) which may, 
therefore, be regarded as the most general form.

Heun's equation was originally constructed as a generalization
of the hypergeometric equation. There are three ways in which Heun's
equation degenerates to hypergeometric equation.

\begin{itemize}

\item
Setting $a = 1$ and $q = \alpha \beta$ yields the canonical hypergeometric
equation for $F(\alpha, \beta. \gamma ; z)$.

\item Setting $\epsilon = 0, q = a \alpha \beta$ also yields the canonical 
hypergeometric equation for $F(\alpha, \beta, \gamma ; z)$.

\item Finally setting $ a = q = 0$ in Eq.(\ref{heuneqn}) yields the hypergeometric
equation for $F(\alpha, \beta, \alpha+\beta - \delta +1 ; z)$.

\end{itemize}

{\it Lam\'e's equation} is a special case of Heun's general equation (Eq.(\ref{heuneqn}))
for which
\baray
\gamma = \delta = \epsilon = \frac{1}{2}~,
\earay
and thus $\alpha + \beta = \frac{1}{2}$. It is common to write
\baray
\alpha = -\frac{1}{2}\nu, ~~\beta = \frac{1}{2} (\nu + 1),~~ q =
-\frac{1}{4}ah~.
\earay
With these redefinitions
the equation takes the canonical form of Lam\'e's equation
\baray
\label{lame}
y''(z) + \frac{1}{2} \left( \frac{1}{z} + \frac{1}{z-1} + \frac{1}{z-a} \right)
y'(z) + \frac{ah - \nu(\nu+1)z}{4z(z-1)(z-a)} y(z) = 0
\earay
The parameter $\nu$ is called the {\it order} of the equation.

Solutions can be constructed to Heun's or Lam\'e's differential equation
both as
a Frobenius series in $z$ and as a series in hypergeometric functions.
In this paper we have made use of the series solution. Upon constructing
a series solution to Heun's equation one typically obtains three-term
recursion relations. It is, in general, impossible to write down the general
solution for three-term recursion relations, unlike two-term
recursion relations. One has to construct the coefficients of the
different powers
order by order.
For further details and discussions on Heun's differential equation
see reference \cite{Heun:1995hn}.

\section{Appendix: The Horizon}
\label{sec:appendix-b-}

When we work with global coordinates
in a de Sitter spacetime, at late times we simply get the usual flat
slice results. So the horizon is $L^{-1}$. Let us find the late time
horizon for the geometry at hand. The event horizon at time $t$ is given
by
\baray
l(t) = a(t) \int_{t}^{\infty}\frac{dt'}{a(t')}
\earay
Using Eq.(\ref{barrel1}) for $a(t)$, we get
\baray
\label{l}
l(t) = \frac{L^{-1}}{\sqrt{1+\Delta^2}}\sqrt{\Delta^2 + \cosh^2(Lt)}
\left( I(\frac{\pi}{2},\delta) - I(\theta, \delta)\right)
\earay
where $\theta = \arcsin(\tanh(L t))$, $\delta = \Delta/\sqrt{1+\Delta^2}$,
and $ I(\alpha, \beta)$ is the
elliptic integral of the first kind defined as
\baray
 I(\alpha, \beta) = \int_{0}^{\alpha}\frac{d\phi}
{\sqrt{1-\beta^2\sin^2\phi}}
\earay

At late times $\theta \to \pi/2$. We shall use
the following identities
\baray
 I(\frac{\pi}{2}, \delta)&=&\frac{\pi}{2} F\left(\frac{1}{2},
\frac{1}{2};1
;\delta^2  \right), \nonumber \\
 I(\theta, \delta)&=&\theta F\left(\frac{1}{2},\frac{1}{2};1
;\delta^2  \right) - \sin \theta \cos \theta \left(a_0 + \frac{2}{3}a_1
\sin^2 \theta + \frac{2.4}{3.5}a_2 \sin^4 \theta + ...\right)
\earay
where $F\left(\frac{1}{2},\frac{1}{2};1;\delta^2  \right)$ is
the hypergeometric series,
and the coefficients are given by
\baray
a_0&=& F\left(\frac{1}{2},\frac{1}{2};1;\delta^2  \right) - 1, \nonumber \\
a_n&=&a_{n-1} - \left( \frac{(2n-1)!!}{2^n n!}\right)^2 \delta^{2n}.
\earay

At late times, $\sqrt{\Delta^2 + \cosh^2(L t)} \to e^{L t}/2$, and
$\pi/2 - \theta  \to 2e^{-L t}$. So we get the following late time
behavior for the horizon
\baray
l(t) &=& \frac{L^{-1}}{\sqrt{1+\Delta^2}} \left( F\left(\frac{1}{2},
\frac{1}{2};1;\delta^2  \right)  + a_0 + \frac{2}{3}a_1 + \frac{2.4}{3.5}
a_2 + ... \right) \nonumber \\
&\simeq& L^{-1}+ \ldots
\earay
where we use the identity $F(1/2,1;1;\delta^2) = 1/\sqrt{1-\delta^2}$
and the fact that $\delta^2 = \Delta^2/(1+\Delta^2)$.
So at late times the horizon is the same as the de Sitter horizon
$L^{-1}$.


\begin{thebibliography}{10}


\bibitem{Peiris:2003ff}
  H.~V.~Peiris {\it et al.},
  ``First year Wilkinson Microwave Anisotropy Probe (WMAP) observations:
  Implications for inflation,''
  Astrophys.\ J.\ Suppl.\  {\bf 148}, 213 (2003)
  [arXiv:astro-ph/0302225].

\bibitem{Berera:1995ie}
  A.~Berera,
  ``Warm Inflation,''
  Phys.\ Rev.\ Lett.\  {\bf 75}, 3218 (1995)
  [arXiv:astro-ph/9509049].
  \\
  L.~M.~H.~Hall, I.~G.~Moss and A.~Berera,
  ``Scalar perturbation spectra from warm inflation,''
  Phys.\ Rev.\ D {\bf 69}, 083525 (2004)
  [arXiv:astro-ph/0305015].


\bibitem{McGreevy:2005ci}
  J.~McGreevy and E.~Silverstein,
  ``The tachyon at the end of the universe,''
  JHEP {\bf 0508}, 090 (2005)
  [arXiv:hep-th/0506130].


\bibitem{Danielsson:2005cc}
  U.~H.~Danielsson,
  ``Inflation as a probe of new physics,''
  JCAP {\bf 0603}, 014 (2006)
  [arXiv:hep-th/0511273].
  \\
  U.~H.~Danielsson,
  ``Transplanckian energy production and slow roll inflation,''
  Phys.\ Rev.\ D {\bf 71}, 023516 (2005)
  [arXiv:hep-th/0411172].

\bibitem{Bouhmadi-Lopez:2002qz}
  M.~Bouhmadi-Lopez, L.~J.~Garay and P.~F.~Gonzalez-Diaz,
  Phys.\ Rev.\ D {\bf 66}, 083504 (2002)
  [arXiv:gr-qc/0204072].


\bibitem{Hartle:1983ai}
J.~B. Hartle and S.~W. Hawking,
`` Wave function of the universe,''  {\em
  Phys. Rev.} {\bf D28} (1983) 2960; \\

\bibitem{Vilenkin:1982de}
A.~Vilenkin,
``Creation of universes from nothing,''  {\em Phys. Lett.} {\bf
  B117} (1982) 25;

\bibitem{Linde:1984mx}
A.~D. Linde,
`` Quantum creation of the inflationary universe,''
{\em Nuovo Cim. Lett.} {\bf 39} (1984) 401; \\


\bibitem{Sarangi:2006eb}
  S.~Sarangi and S.~H.~Tye,
  ``A Note on the Quantum Creation of Universes,''
  arXiv:hep-th/0603237.


\bibitem{Sarangi:2005cs}
  S.~Sarangi and S.~H.~Tye,
  ``The boundedness of Euclidean gravity and the wavefunction of the
  universe,''
  arXiv:hep-th/0505104.


\bibitem{Brustein:2005yn}
  R.~Brustein and S.~P.~de Alwis,
  ``The landscape of string theory and the wave function of the universe,''
  Phys.\ Rev.\ D {\bf 73}, 046009 (2006)
  [arXiv:hep-th/0511093].

\bibitem{Barvinsky:2006uh}
  A.~O.~Barvinsky and A.~Y.~Kamenshchik,
  ``Cosmological landscape from nothing: Some like it hot,''
  arXiv:hep-th/0605132.

\bibitem{Vilenkin:1998rp}
A.~Vilenkin,
`` The quantum cosmology debate,''
gr-qc/9812027.

\bibitem{Mottola:1984ar}
  E.~Mottola,
  ``Particle Creation In De Sitter Space,''
  Phys.\ Rev.\ D {\bf 31}, 754 (1985).


\bibitem{Allen:1985ux}
  B.~Allen,
  ``Vacuum States In De Sitter Space,''
  Phys.\ Rev.\ D {\bf 32}, 3136 (1985).


\bibitem{Allen:1987tz}
  B.~Allen and A.~Folacci,
  ``The Massless Minimally Coupled Scalar Field In De Sitter Space,''
  Phys.\ Rev.\ D {\bf 35}, 3771 (1987).


\bibitem{Vilenkin:1982wt}
  A.~Vilenkin and L.~H.~Ford,
  ``Gravitational Effects Upon Cosmological Phase Transitions,''
  Phys.\ Rev.\ D {\bf 26}, 1231 (1982).


\bibitem{Leblond}
  F.~Leblond, D.~Marolf and R.~C.~Myers,
  JHEP {\bf 0206}, 052 (2002)
  [arXiv:hep-th/0202094].

\bibitem{Starobinsky:1996ek}
  A.~A.~Starobinsky,
  ``Spectrum of initial perturbations in open and closed inflationary models,''
  arXiv:astro-ph/9603075.

\bibitem{Chernikov:1968zm}
  N.~A.~Chernikov and E.~A.~Tagirov,
  ``Quantum theory of scalar fields in de Sitter space-time,''
  Annales Poincare Phys.\ Theor.\ A {\bf 9}, 109 (1968).

\bibitem{Heun:1995hn}
  Ronveaux, A. (Ed.).
  Heun's Differential Equations. Oxford, England: Oxford University
  Press, 1995.

\bibitem{Strominger} 
  R.~Bousso, A.~Maloney and A.~Strominger,
  ``Conformal vacua and entropy in de Sitter space,''
  Phys.\ Rev.\ D {\bf 65}, 104039 (2002)
  [arXiv:hep-th/0112218].


\bibitem{Halliwell:1984eu}
  J.~J.~Halliwell and S.~W.~Hawking,
  ``The Origin Of Structure In The Universe,''
  Phys.\ Rev.\ D {\bf 31}, 1777 (1985).

\bibitem{Vachaspati:1989wf}
  T.~Vachaspati,
  ``De Sitter invariant states from Quantum Cosmology'',
  Phys.\ Lett.\ B {\bf 217}, 228 (1989).

\bibitem{GR2000}
Gradshteyn and Ryzhik's Table of Integrals, Series, and Products.
Alan Jeffrey and Daniel Zwillinger (eds.)
Sixth edition (July 2000)

\bibitem{Firouzjahi:2004mx}
  H.~Firouzjahi, S.~Sarangi and S.~H.~H.~Tye,
  ``Spontaneous creation of inflationary universes and the cosmic  landscape,''
  JHEP {\bf 0409}, 060 (2004)
  [arXiv:hep-th/0406107].


\bibitem{Aazami:2005jf}
  A.~Aazami and R.~Easther,
  ``Cosmology from random multifield potentials,''
  JCAP {\bf 0603}, 013 (2006)
  [arXiv:hep-th/0512050].


\bibitem{Arkani-Hamed:2005yv}
  N.~Arkani-Hamed, S.~Dimopoulos and S.~Kachru,
  ``Predictive landscapes and new physics at a TeV,''
  arXiv:hep-th/0501082.


\bibitem{Watson:2006px}
  S.~Watson, M.~J.~Perry, G.~L.~Kane and F.~C.~Adams,
  ``Inflation without inflaton(s),''
  [arXiv:hep-th/0610054].

\bibitem{Greene:2005aj}
  B.~Greene, K.~Schalm, J.~P.~van der Schaar and G.~Shiu,
  ``Extracting new physics from the CMB,''
  eConf {\bf C041213}, 0001 (2004)
  [arXiv:astro-ph/0503458].

\end{thebibliography}
\end{document}